\documentclass[12pt,onecolumn,draftclsnofoot]{IEEEtran}
\usepackage{times}
\usepackage[final]{graphicx}
\usepackage{amsmath,amsfonts,mathrsfs}
\usepackage{psfrag}
\usepackage{amssymb,amsbsy}
\usepackage{times}

\usepackage{epsf} 
\usepackage{cite}

\newcommand\ignore[1]{}
\begin{document} 

\title{To Code or Not to Code Across Time: Space-Time Coding with Feedback}
\author{ 
{\large{Che Lin, Vasanthan Raghavan, and Venugopal V. Veeravalli$^{*}$}} 
\thanks{The authors are with the Coordinated Science Laboratory and the 
Department of Electrical and Computer Engineering, 
University of Illinois at Urbana-Champaign, Urbana, IL 61801 USA. 
Email: {\tt{\{clin, vvv\}@uiuc.edu, vasanthan\_raghavan@ieee.org.}} 
$^*$Corresponding author.} 
\thanks{This research was supported in part by the NSF award \#CCF 0431088 
through the University of Illinois, and by a Vodafone Foundation Graduate 
Fellowship. This paper was presented in part at the IEEE Global Communications 
Conference, Washington, DC, 2007.} 
}
\maketitle

\newtheorem{thm}{Theorem}
\newtheorem{lem}{Lemma}
\newtheorem{example}{Example}
\newtheorem{defn}{Definition}
\newtheorem{prop}{Proposition}
\newtheorem{remark}{Remark}
\newtheorem{assumption}{Assumption}
\newtheorem{property}{Property}
\newtheorem{conjecture}{Conjecture}
\newtheorem{algm}{Algorithm}

\newcommand{\note}{{\bf Note: }}
\newcommand{\beq}{\begin{equation}}
\newcommand{\eeq}{\end{equation}}
\newcommand{\beqa}{\begin{eqnarray}}
\newcommand{\eeqa}{\end{eqnarray}}
\newcommand{\mb}{\mathbf}
\newcommand{\mt}{\textrm}
\newcommand{\bnum}{\begin{enumerate}}
\newcommand{\enum}{\end{enumerate}}
\newcommand{\bu}{ {\bf u}} 

\newcommand{\tH}{\widetilde{{ \rm H}\phi}}
\newcommand{\tn}{\widetilde{n}}
\renewcommand{\th}{\widetilde{h}}
\newcommand{\tA}{\widetilde{{A}}}
\newcommand{\tY}{\widetilde{{Y}}}
\newcommand{\tW}{\widetilde{{W}}}
\newcommand{\tmH}{\widetilde{ \mb{H}}}
\newcommand{\tmW}{\widetilde{\mb{W}}}
\newcommand{\tmY}{\widetilde{\mb{Y}}}

\newcommand{\E}{\mathbb{E}}
\newcommand{\R}{\mathbb{R}}
\newcommand{\C}{\mathbb{C}}
\newcommand{\N}{\mathbb{N}}
\newcommand{\HH}{\mathbb{H}}
\renewcommand{\S}{\mathbb{S}}
\newcommand{\bQ} { {\bf Q} }

\newcommand{\sfh}{ {\sf{H}}}
\newcommand{\bo} { {\mathbf{0}} }
\newcommand{\trace}{ {\mathrm{Tr}} }
\newcommand{\diag}{ \mathrm{diag} }

\newcommand{\Y}{\tilde{\mb{Y}}}
\newcommand{\bA} { {\mathbf{A}} }
\renewcommand{\H}{\tilde{\mb{H}}}
\renewcommand{\N}{\tilde{\mb{N}}}
\newcommand{\A}{\widetilde{A}}
\newcommand{\Ad}{\widetilde{A}\widetilde{A}^H}
\renewcommand{\b}{\mb{b}}

\newcommand{\pf}{{\bf Proof: }}

\newcommand{\bmY}  {{\mathbf{Y}} }
\newcommand{\bmX}  {{\mathbf{X}} }
\newcommand{\bmH}  {{\mathbf{H}} }
\newcommand{\bmW}  {{\mathbf{W}} }
\newcommand{\bmN}  {{\mathbf{N}} }

\newcommand{\bmx}  {{\mathbf{x}} } 
\newcommand{\bms}   {{\mathbf{s}} }
\newcommand{\bmn}   {{\mathbf{n}} }
\newcommand{\bmy}   {{\mathbf{y}} }
\newcommand{\bmh}   {{\mathbf{h}} }
\newcommand{\bmv}   {{\mathbf{v}} } 
\newcommand{\bma}   {{\mathbf{a}} } 
\newcommand{\snr}   {{\sf{SNR}}} 

\newcommand{\bLambda}  {{\mathrm{\Lambda}} } 
\newcommand{\bfLambda}  {{\mathbf{\Lambda}} } 
\newcommand{\bU}  {{\mathbf{U}} }
\newcommand{\bH}  {{\mathbf{H}} }
\newcommand{\bEe}{{\mathbf{E}}}    
\newcommand{\iid} {  {\sf{iid}} }
\newcommand{\ind} { {\sf{ind}} }

\newcommand{\mse}{ {\sf mse} }
\newcommand{\mmse}{ {\sf{mmse}} }
\newcommand{\bflambda}{ {\mathbf{\Lambda}} }
\newcommand{\bV} { {\mathbf{V}} }
\newcommand{\bv} { {\mathbf{v}} }
\newcommand{\bx} { {\mathbf{x}} }
\newcommand{\bP} { {\mathbf{P}} }
\newcommand{\bs} { {\mathbf{s}} }
\newcommand{\bS} { {\mathbf{S}} }
\newcommand{\ba} { {\mathbf{a}} }
\newcommand{\bb} { {\mathbf{b}} }
\newcommand{\bz} { {\mathbf{z}} }
\newcommand{\bZ} { {\mathbf{Z}} }
\newcommand{\bAt} { {\widetilde{\bA }} }
\newcommand{\stat} { {\sf stat} }
\newcommand{\bI}{{\bf I}}
\newcommand{\dg}{\dagger}

\newcommand{\vsp}{\vspace{0.1in} }
\newcommand{\hsp}{\hspace{0.1in} }
\newcommand{\vspp}{\vspace{0.05in} }
\newcommand{\hspp}{\hspace{0.05in} }
\newcommand{\hsppp}{\hspace{0.02in} }
\newcommand{\vsppp}{\vspace{0.02in} }

\newcommand{\vspn}{\vspace{-0.1in} }
\newcommand{\vspnn}{\vspace{-0.05in} }
\newcommand{\hspn}{\hspace{-0.1in} }
\newcommand{\hspnn}{\hspace{-0.05in} }

\newcommand{\perf}{ {\sf{perf}}  } 
\newcommand{\bef} { {\sf{bf}} } 
\newcommand{\st} {{\sf{st}}} 

\newcommand{\bX} { {\bf X} }
\newcommand{\ord}{{\mathcal{O}}}
\newcommand{\cb}{{\mathcal{C}}}
\newcommand{\cbt}{{\mathcal{ \widetilde{C} }}}
\newcommand{\littleo}{{\mathnormal{o}}}
\newcommand{\opt}{ {\sf opt} }
\newcommand{\bY}{ {\bf Y} } 

\newcommand{\gammarone} { \mu_{r,\hsppp 1}  } 
\newcommand{\gammartwo} {  \mu_{r, \hsppp 2} }
\newcommand{\gammatone} { {\mathrm{Gap}}_{t,\hsppp 12} }
\newcommand{\gammattwo} { {\mathrm{Gap}}_t }
\newcommand{\gammatc} { {\mathrm{Gap}}_t^c }
\newcommand{\gammarc} { \mu_{r, \hsppp 2}^{ c} }
\newcommand{\gammarcone}  {  \mu_{ r,\hsppp 2 }^{c, \hsppp ub}  }

\newcommand{\spatpow}{{\bf P}_{\sf spat}} 
\newcommand{\temppow}{ {\bf P}_{\sf temp} } 
\newcommand{\symbpow}{ {\bf P}_{\sf symb} }

\begin{abstract} 
\noindent 
Space-time codes leverage the availability of multiple antennas to enhance the 
reliability of communication over wireless channels. While space-time codes have 
initially been designed with a focus on open-loop systems, recent technological 
advances have enabled the possibility of low-rate feedback from the receiver to 
the transmitter. The focus of this paper is on the implications of this feedback in 
a single-user multi-antenna system with a general model for spatial correlation. 
We assume a limited feedback model, that is, a coherent receiver and statistics along 
with $B$ bits of quantized channel information at the transmitter. We study space-time 
coding with a family of {\em linear dispersion (LD) codes} that meet an 
additional orthogonality constraint so as to ensure low-complexity decoding. Our results 
show that, when the number of bits of feedback ($B$) is small, a space-time coding 
scheme that is equivalent to beamforming and {\em does not} code across time is optimal 
in a weak sense in that it maximizes the average received $\snr$. As $B$ increases, this 
weak optimality transitions to optimality in a strong sense which is characterized 
by the maximization of the average mutual information. Thus, from a system designer's 
perspective, our work suggests that beamforming may not only be attractive from a 
low-complexity viewpoint, but also from an information-theoretic viewpoint. 

\end{abstract}


\section{Introduction} 
\label{sec:intro} 
Multipath fading, where the strength of the received signal fluctuates randomly, 
results in errors in signal propagation. These errors can be combated in practice 
by using multi-antenna diversity techniques at the transmitter and the receiver. 
The focus of this work is on the reliability aspect of multi-input multi-output 
(MIMO) systems under certain assumptions on communication models motivated by 
wireless systems in practice. In particular, we assume a block fading, narrowband 
model for the channel variation in time and frequency, and focus on 
realistic models for spatial correlation and channel state information (CSI) at the 
transmitter and the receiver. 

In this setting, the low-complexity beamforming scheme has attracted significant 
theoretical attention and has been studied with both perfect CSI at the 
transmitter~\cite{paulraj_book}, as well as with partial 
channel knowledge at the transmitter~\cite{david_grass,kiran_outage,vasanth_limfb_bf}. 
On the other hand, initial works on space-time codes assume no CSI at the transmitter 
and study the reliability of information transmission with uncoded inputs, that is, 
the input 
symbols are independent from one coherence block to another. Reliability can be 
improved by using certain {\em delay diversity techniques}~\cite{delay_div} and 
these schemes can be extended to a more general framework, collectively known as 
{\em space-time trellis codes}~\cite{sttc}. Though space-time trellis codes are 
near-optimal in the MIMO setting, they suffer from decoding complexity that is 
exponential with the rate or the number of transmit antennas. To overcome these 
difficulties, {\em orthogonal space-time block codes} (OSTBC)~\cite{alamouti,orth_designs} 
that achieve the full diversity order\footnote{The diversity order of a 
code is defined as the exponent of the rate at which error probability decays with $\snr$.} 
of multi-antenna systems and offer the additional advantage of low decoding complexity 
have been proposed. 

Maximizing the diversity order, while being a useful design criterion, is only 
applicable for uncoded transmissions. In practice, a space-time code is used as an 
{\em inner code} in concatenation with an error correction code that is 
used as an {\em outer code} and is designed\footnote{For example, recent works on 
low-density parity check codes (see, e.g.,~\cite{tenbrink_ldpc}) have 
shown that it is possible to construct outer codes that come close to achieving the 
mutual information between the input and the output of the inner space-time code.} to 
achieve maximum possible rate for a given $\snr$. Clearly, in this setting, diversity 
order is not of great importance because the outer code can exploit the full diversity 
benefit of the channel by coding across different space-time coding blocks. Since the 
outer code helps the concatenated transmitter in approaching capacity, the mutual 
information is a meaningful design metric for the inner space-time block codes if 
soft decisions are allowed at the space-time decoder. In the no CSI case, it has been 
established that~\cite{yibo_jiang,bresler_hajek} OSTBC are also optimal from a mutual 
information viewpoint. 

The assumption of no CSI at the transmitter is too pessimistic and does not 
capture reality where either statistical or partial channel 
knowledge~\cite{limfb_honig} at the transmitter is possible. In this context, 
when only the statistics are available at the transmitter, it is known that OSTBC are no 
longer optimal within the class of linear space-time codes~\cite{clin_ldcorr}. 
In the more practical limited feedback case, there have been some recent 
works~\cite{jongren_ostbc,larsson_ostbc,david_ostbc,ekbatani,akhtar} 
on the design of space-time codes with CSI feedback. However, much of this body of 
work ignores spatial correlation and focuses on weighted OSTBC which result in a 
very restrictive set of linear operations on the inner space-time codes. As witnessed 
in the statistical case~\cite{clin_ldcorr}, spatial correlation can potentially lead 
to significant changes in code design criterion and optimal signaling. {\em Thus, 
our goal in this work is to address optimal signaling with partial channel 
knowledge at the transmitter and more generally, whether coding across time is 
necessary in partial CSI systems.} 

The general framework of {\em linear dispersion (LD) codes}, introduced 
in~\cite{hassibi_ld}, subsumes all linear space-time codes and hence provides a 
natural framework for studying both beamforming as well as space-time code design 
in a unified way given that partial CSI is available at the transmitter. In an LD code, 
each symbol that is transmitted over the channel is some linear combination 
of the inputs and their complex conjugates and the codes are designed to maximize 
the mutual information between the input and the output of the space-time code. 
While the generality of the LD framework leads to some complications in code design, 
recent works~\cite{heath_ld,wang_ld,akbar_gcom} show that systematic 
LD code constructions are still possible. 

In this work, we impose an additional {\em Generalized Orthogonal Constraint 
(GOC)}~\cite{orth_designs,clin_ldcorr} on the LD codes so that they enjoy the same 
low-complexity of decoding\footnote{Satisfaction of the GOC ensures that the joint 
maximum-likelihood (ML) decoding of the vector input reduces to individual 
ML decoding of the scalar inputs.} as OSTBC. That is, we consider the set of 
{\em orthogonal LD} codes. The search for the optimal orthogonal LD codes provides 
significant insights to answer whether coding across time is necessary or not. 

\noindent {\bf{\emph{Contributions:}}}
\begin{itemize}
\item 
We first show that when there is perfect CSI at the transmitter, the optimal 
power allocation across the different input symbols is uniform. Furthermore, the 
rank of the optimal LD code 
is one and since rank-one LD codes are equivalent to beamforming, the {\em 
optimal perfect CSI scheme does not code across time.} 

\item 
When only statistical information is available at the transmitter, we establish that 
uniform symbol power allocation is still optimal. It is also known from~\cite{clin_ldcorr} 
that the rank of the optimal linear space-time code is in general 
dependent on $\snr$ and channel correlation. Furthermore, for any correlation, the 
rank is a non-decreasing function of $\snr$. Thus the {\em optimal statistical 
scheme codes across time, in general.} 

\item 
In the partial CSI case, we first establish the optimality of uniform symbol power 
allocation, irrespective of the level of channel knowledge at the transmitter. 
On the question of spatial power allocation, it is natural to expect a smooth transition 
for the rank of the optimal scheme as the quality of channel information at the 
transmitter gets successively refined (that is, as the number of bits of feedback $B$ 
increases). {\em Surprisingly, we show that rank-one schemes enjoy strong 
optimality properties.} When $B$ is sufficiently large (for example, 
$B \gg \log(N_t)$ with $N_t$ 
denoting the transmit antenna dimension), we show that rank-one schemes maximize 
the average mutual information, while in contrast when $B$ is small, we show a slightly 
weaker result: Rank-one schemes maximize the average received $\snr$. While we expect 
a transition from strong optimality to weak optimality for small values of $B$ (as a 
function of the $\snr$ and the spatial correlation), numerical studies suggest that for 
most practical correlation, this $\snr$ is too large from a practical standpoint. 
{\em Thus, our results suggest that the optimal scheme under the orthogonal LD code 
framework and quantized feedback corresponds to not coding across time}. 


\item 
The optimality of rank-one schemes (beamforming) implies that the low-complexity 
advantage of scalar coding is justified from an information theoretic sense. 


\end{itemize}

\noindent {\bf{\emph{Notations:}}} 
We use $\bX(i,j)$ and $\bX(i)$ to denote the $i,j$-th and $i$-th 
diagonal entries of a matrix $\bX$. The 
conjugate transpose and regular transpose 
are denoted by 
$(\cdot)^{\dg}$ and $(\cdot)^{T}$ 
while $\bEe [\cdot]$ and $\trace(\cdot)$ 
stand for the expectation and the trace 
operators, respectively. 
We say that a singular value decomposition of a matrix is in its 
{\em standard ordering} if its singular values are arranged in non-increasing order. 
Further, if the matrix is Hermitian, $\lambda_{\max}(\cdot)$ 
denotes the largest 
eigenvalue. 


\section{System Setup}
\label{sec2} 
We consider a single-user MIMO communication system with $N_t$ transmit and $N_r$ 
receive antennas. The multi-antenna channel matrix experiences fading in time, 
frequency, and space. In this paper, we assume a narrowband, block fading model 
for the channel. That is, the channel is frequency flat and remains 
constant across a block of length $N_c$ symbols and fades ergodically from block 
to block. With these simple models for the evolution of the channel across time 
and frequency, the main focus is on the spatial aspect. 

To overcome the impediments of fading, we will consider the design of space-time 
codes and view the channel across the block length $N_c$ as corresponding to 
one {\emph{channel use}}. The discrete-time, complex baseband model under this 
setting is given by 
\beq 
\nonumber 
\bmY = \sqrt{\frac{\rho}{N_t}} \bmH \bmX + \bmW
\eeq 
where $\bmX \in \C^{N_t \times N_c}$ is the transmitted signal matrix, 
$\bmY \in \C^{N_r\times N_c}$ is the received signal matrix, 
$\bmH \in \C^{N_r\times N_t}$ corresponds to the channel matrix, and 
$\bmW \in \C^{N_r\times N_c}$ denotes the complex additive white Gaussian noise 
with i.i.d.\ entries, $\bmW(i,j)\sim {\cal CN}(0,1)$. We assume an average power 
constraint on $\bmX$ given by $\bEe \left[ \trace \left( \bmX \bmX^{\dg} \right) 
\right] \leq N_t N_c$ that results in a transmit power constraint $\rho$ over 
each symbol duration. 

\subsection{Spatial Correlation} 
We now describe the spatial fading framework used in this work. It has been 
well-documented that the assumption of zero-mean Rayleigh fading is an accurate 
model for $\bH$ in a non line-of-sight setting. Thus the complete channel 
statistics are described by the second-order moments. Rich scattering environments 
are accurately modeled by the commonly used i.i.d.\ model where the channel entries 
are i.i.d.\ ${\cal{CN}}(0,1)$. However, the i.i.d.\ model is not accurate in 
describing realistic propagation environments. Various statistical models have been 
proposed to overcome the deficiencies associated with the i.i.d. model. 

The most general, mathematically tractable spatial correlation model is based on a 
decomposition of the channel onto its canonical coordinates: the eigen-bases of the 
transmit and the receive covariance 
matrices~\cite{canonical_jayesh,canonical_bonek,tulino_ind}. 
The canonical model assumes that the auto- and the cross-correlation matrices on 
both the transmitter and the receiver sides have the same eigen-bases, and 
exploits this redundancy to decompose ${\bf H}$ as 
\begin{eqnarray}
\label{canl}
\bH = \bU_r \hsppp \bH_{  \ind } \hsppp  \bU_t^{\dg}
\end{eqnarray} 
where $\bH_{\ind}$ has independent, but not necessarily identically distributed 
entries. $\bU_r$ and $\bU_t$ are eigenvector matrices corresponding to the receive 
and the transmit covariance matrices which are defined as ${\bf \Sigma}_r = 
\bEe[ \bH \bH^{\dg} ]$ and ${\bf \Sigma}_t = \bEe[ \bH^{\dg}\bH ]$, respectively. 
It can be checked that~\cite{canonical_jayesh} the model in~(\ref{canl}) reduces 
to some well-known models 
like the separable correlation model or the virtual representation 
framework~\cite{akbar,akbar_and_venu,venu_capacity}.

\ignore{ 
It can be checked that the model in (\ref{canl}) reduces to some well-known 
spatial correlation models: 
\begin{itemize}
\item 
When $\bH_{  \ind }$ is assumed to have the form $\bfLambda_r^{1/2} \hsppp 
\bH_{\iid} \hsppp \bfLambda_t^{1/2}$, the canonical model reduces to the 
separable correlation framework~\cite{chuah}. 
\item 
When uniform linear arrays (ULAs) of antennas are used at the transmitter and the 
receiver, $\bU_r$ and $\bU_t$ are well-approximated by discrete Fourier transform 
matrices and the canonical model reduces to the virtual representation
framework~\cite{akbar,akbar_and_venu,venu_capacity}. In contrast to the general 
model in~(\ref{canl}), the virtual representation offers many attractive properties: 
1) The matrices $\bU_r$ and $\bU_t$ are {\em fixed} and independent of the 
underlying scattering environment and the spatial eigenfunctions are beams in the 
virtual directions. Thus, the virtual representation is physically more intuitive 
than~(\ref{canl}), 2) It is only necessary that the entries of $\bH_{\ind}$ be 
independent, but not necessarily Gaussian, a criterion important as antenna 
dimensions increase, and 3) The case of specular (or line-of-sight) scattering can 
be easily incorporated with the virtual representation framework. 
\end{itemize}
The readers are referred 
to~\cite{canonical_jayesh,canonical_bonek,akbar,akbar_and_venu,venu_capacity} for more 
details on the channel models, the underlying assumptions on the scattering 
environment, their implications, and connections to previous work in modeling. 
}

\subsection{Channel State Information}
We will assume perfect CSI at the receiver in this work. However, perfect CSI at the 
transmitter is {\em not} feasible due to fast time variations of the channel that 
leads to a high cost associated with channel feedback$/$reverse-link 
training\footnote{In case of Time-Division Duplexed (TDD) systems, the reciprocity 
of the forward and the reverse links can be exploited to train the channel on the 
reverse link. In case of Frequency-Division Duplexed (FDD) 
systems, the channel information acquired at the receiver has to be fed back.}. 
Nevertheless, we assume that the channel statistics, which change much more slowly 
than the channel realizations, 
can be learned reliably at both the ends. In practice, besides the statistical 
information, there is usually a viable low rate feedback link from the receiver to 
the transmitter. 
Thus recent attention, in both theory and practice, has shifted towards understanding 
the implications of partial CSI at the transmitter (most notably, in the form of 
limited or quantized feedback~\cite{limfb_honig}) on the performance of communication 
systems. In this work, we assume an error-free, negligible-delay limited feedback 
link where $B$ bits of channel information are conveyed per channel use. 

\subsection{Signaling Scheme - Linear Dispersion Codes} 
As mentioned in the introduction, the coding problem for the MIMO channel $\bH$ 
can be separated into the design of an inner space-time block code and an 
outer code. 
Accordingly, input data $\bmx[t]$ 
is demultiplexed into $K$ data-streams denoted by $\bmx_1[t], \cdots, \bmx_{K}[t]$ 
for the space-time encoder at a given symbol time $t$. We make the following 
simplifying assumptions on the input symbols in this work. 
\newline 
\noindent {\bf{\emph{Assumption 1:}}} 
\begin{itemize} 
\item 
The data-streams corresponding to $\bmx_k[t]$ are i.i.d.\ across time
for all $k$ and they are drawn from some {\emph{real}} constellation with marginal 
distribution $p(x_k)$. The mean of $\bmx_k[t]$ is zero. 
\item
For any $t$ and all $i,j$ such that $i \neq j$, $\bmx_i[t]$ and
$\bmx_j[t]$ are independent. 
\end{itemize}
The second assumption can be justified if $\bmx_1[t],\hsppp \cdots, \hsppp 
\bmx_{K}[t]$ are 
produced as outputs of independent scalar outer encoders as in the V-BLAST signaling 
scheme. Applications involving the use of bit-interleaved codes at the outer encoder 
also justify the second assumption.
Furthermore, both assumptions can be justified if the data coming from the encoder is 
fed through a random interleaver, a very practical assumption.
Since $\bmx_k[t]$ are i.i.d.\ across time, we will drop the time index $t$
in the ensuing discussion. 

While arbitrarily structured space-time coding schemes can be considered for 
signaling, in this work, we will focus on a specific LD code-based 
signaling~\cite{hassibi_ld}. The definition of an LD code involves a set of 
dispersion matrices $\{ \bA_k \} \in \C^{N_t \times N_c}$ such that the space-time 
code $\bmX$ is 
\beq 
\label{LD_code} 
\bmX=\sum_{k=1}^K  \bA_k \hsppp \bmx_k
\eeq 
where the symbols $\{\bmx_k\}_{k=1}^K$ satisfy Assumption 1. 
That is, at a given symbol time, the outer encoder produces a set of independent 
symbols $\{\bmx_k\}$ which is then spread across 
the spatial and temporal dimensions through $\{\bA_k\}$. 

It is important to note that LD codes encompass all possible linear space-time codes. 
In addition, we assume that the class of LD codes satisfy the 
{\em Generalized Orthogonal Constraint (GOC)}, that is,  
${\bA}_k  \bA_j^{\dg} + \bA_j \bA_k^{\dg} = \bo$ for all $k,j,\hsppp  k \neq j$. 
It has been shown in~\cite{yibo_jiang,orth_designs} that the GOC is equivalent to 
$p( \bx_1,\ldots,\bx_K|{\bmY}, \hsppp \bH_\ind) = \prod_{k=1}^K 
p(\bx_k| \bmY, \hsppp \bH_\ind)$. That is, the likelihood function factors and 
the complexity of the LD decoder is greatly reduced since the joint 
ML decoding reduces to individual 
ML decoding of each symbol. In other words, the channel decouples 
into $K$ parallel sub-channels. It is important to note that the decoding 
complexity for this class of LD codes, labeled henceforth as {\em orthogonal LD 
codes}, is the same as that achieved by OSTBC. 

After normalizing 
$\bmx_k$ such that $\bEe \left [\bmx_k^2 \right] = 1$, the power constraint is 
applied to $\bA_k$ resulting in $\sum_{k=1}^K \trace ( \bA_k \bA_k^{\dg} ) \leq 
N_t N_c$. 
The power allocated to the $k$-th symbol is $\trace({\bf A}_k {\bf A}_k^{\dg} )$. 
\ignore{ 
The symbol-wise power allocation vector is defined as 
\begin{eqnarray}
\symbpow & =  & \left( {\bf P}_{ {\sf symb}, \hsppp k } \right), {\hspace{0.1in}} 
{\bf P}_{ {\sf symb}, \hsppp k } = \sum_{i,j} | {\bf A}_{k}(i,j) |^2
= \trace({\bf A}_k {\bf A}_k^{\dg} ). \nonumber  
\end{eqnarray}
}

\section{Optimal Signaling Schemes} 
\label{sec_opt_sig} 
In this section, we 
study the problem of optimal LD code construction under different assumptions 
on the available CSI at the transmitter. There are four relevant cases of CSI: 
two extreme cases of no$/$perfect CSI at the transmitter, a reasonable assumption 
where only statistical information is available, and a case where partial CSI in the 
form of quantized feedback is available at the transmitter. 
The first three cases are the subject of this section, and Sec.~\ref{sec:quan_csi} 
deals with the last case in more detail. 

\subsection{No CSI at the Transmitter} 
When no channel information is available at the transmitter, the optimal scheme 
is to assume that the channel is i.i.d. Thus, any space-time code tailored to the 
i.i.d.\ case can be used. In particular, Hassibi and Hochwald~\cite{hassibi_ld} 
have applied the mutual information criterion to design optimal codes (within the 
class of LD codes) with i.i.d.\ Gaussian inputs. Jiang~\cite{yibo_jiang} 
has studied the design of optimal LD codes for i.i.d.\ channels with binary 
inputs and conjectured that the optimal code is the generalized orthogonal design 
introduced in~\cite{orth_designs}. Bresler and Hajek~\cite{bresler_hajek} 
proved the above conjecture and extended the work to arbitrary real inputs. 
The following sections demonstrate how channel information can help improve performance. 

\subsection{Statistical Information at the Transmitter} 
\label{sec_stat}
We now study the structure of the optimal LD codes 
when only statistical information is available at the transmitter. We build on the 
recent work in~\cite{clin_ldcorr} where optimal LD codes are constructed by maximizing 
the average mutual information between the input and the output of the inner code. 
If channel correlation is modeled with the canonical 
framework as in (\ref{canl}), we obtain the following equivalent channel model:
\beq 
\label{virtual_ch_model}
\widetilde{\bmY} = \sqrt{\frac{\rho}{N_t}} \hspp 
\sum_{k=1}^K  {\bH_{\ind}} \hsppp {\widetilde{\bA}}_k \hsppp \bmx_k + 
\widetilde{\bmW}
\eeq 
where ${\widetilde{\bA}}_k = \bU_t^{\dg} \bA_k$, $\widetilde{\bmY} = 
\bU_r^{\dg} \bmY$, and $\widetilde{\bmW} =  \bU_r^{\dg} \bmW$. 
The GOC is equivalent to ${\widetilde{\bA}}_k  {\widetilde{\bA}}_j^{\dg} + 
{\widetilde{\bA}}_j {\widetilde{\bA}}_k^{\dg} = \bo$ for all $k,j,\hsppp  
k \neq j$ and the original power constraint is equivalent to 
$\sum_{k=1}^K \trace \big( {\widetilde{\bA}}_k {\widetilde{\bA}}_k^{\dg} \big) 
\leq N_t N_c$. 
We have the following theorem characterizing the structure of optimal LD codes. 
\begin{thm} 
\label{thm_stat_opt} 
Let $\widetilde{\bmX} = {\bf U}_t^{\dg} \bmX$ 
be an LD code as in~(\ref{LD_code}) with $K$ symbols and 
let the corresponding dispersion matrices be $\{ {\widetilde{\bA}}_k, \hsppp 
k = 1 , \cdots, K \}$. Also, let the input symbols $\bmx_1, \cdots, \bmx_K$ 
satisfy Assumption~1 and the dispersion matrices satisfy the GOC. 
If there exists an LD code satisfying 
the {\emph{power constraint condition}}: 
${\widetilde{\bA}} _k {\widetilde{\bA}}_k^{\dg} = \bfLambda_{\stat}$ for all 
$k$ where $\bfLambda_{\stat}$ is a positive semidefinite diagonal matrix with 
\begin{eqnarray}
\bfLambda_{\stat} = \arg \max_{\bfLambda} \bEe \left[ \varphi \left(\frac{\rho}{N_t} 
\trace \left(  \bH_{\ind} \bfLambda \bH_{\ind}^{\dg} \right) 
\right) \right] {\hspace{0.07in}} {\rm s.t.} {\hspace{0.05in}} 
\trace \left( \bfLambda \right) = \frac{N_t N_c}{ K}, 
\nonumber \end{eqnarray}
then such a code maximizes the average mutual information 
$\bEe \left[ I( {\bmX}; {\bmY}| {\bmH} = \sfh) \right]$ 
and achieves ergodic capacity. 
\end{thm}
{\vspace{0.05in}}
\begin{proof} 
See Appendix~\ref{app_2}. 
\end{proof}

Theorem \ref{thm_stat_opt} states that uniform symbol power allocation across the 
data-streams is optimal from an average mutual information viewpoint. The optimal 
spatial power allocation is given by $\bfLambda_{\stat}$, and in general, 
$\bfLambda_{\stat}$ excites multiple modes non-uniformly. We now elaborate 
on the structure of $\{ \widetilde{\bA}_k \}$. 
Given that $r$ denotes the number of spatial modes excited by the optimal statistical 
scheme, it is straightforward to check that $N_c \geq r$ is necessary. This 
follows by assuming a generic singular value decomposition for $\widetilde{\bA}_k$ 
and checking that the power constraint condition holds. Furthermore, it can be seen 
that 
\begin{eqnarray}
\widetilde{\bA}_k &= & \left\{ 
\begin{array}{cc}
\left[ \sqrt{ \bfLambda_{\stat}} \hsp {\bf 0}_{N_t \times N_c - N_t}  \right] 
\bY_k^{\dg} & {\rm if} \hspp N_c \geq N_t \\ 
\left[ \sqrt{\bfLambda_{\stat}} \hsppp \right]_{\sf prin} \bY_k^{\dg} & 
{\rm if} \hspp r \leq N_c < N_t 
\end{array} \right. \nonumber 
\end{eqnarray}
where $\bY_k$ is an arbitrary $N_c \times N_c$ unitary matrix and 
$\left[ \sqrt{\bfLambda_{\stat}} \hsppp \right]_{\sf prin}$ is the $N_t \times N_c$ 
principal sub-matrix of $\sqrt{\bfLambda_{\stat}}$. 
With the above structure for $\widetilde{\bA}_k$ and with $\widetilde{\bY}_k$ denoting 
the $N_c \times r$ principal sub-matrix of $\bY_k$, we need 
\begin{eqnarray}
\widetilde{\bY}_k^{\dg} \widetilde{\bY}_j + \widetilde{\bY}_j^{\dg} \widetilde{\bY}_k = 
{\bf 0} \hspp \hsppp {\rm for} \hspp {\rm all} \hspp k \neq j 
\label{connnn}
\end{eqnarray}
to meet the GOC. If $rK \leq N_c$,~(\ref{connnn}) can be met by letting 
$\{ \widetilde{\bY}_k \}$ to be a set of $r$ distinct columns of a random 
$N_c \times N_c$ unitary matrix. In fact, this choice leads to a stronger condition 
where $\widetilde{\bY}_k^{\dg} \widetilde{\bY}_j = {\bf 0}$ for any $k \neq j$. 
Initial studies suggest that $r K \leq 2 N_c$ is both necessary and sufficient 
for a feasible construction that meets~(\ref{connnn}). These results and their 
connections to constructions via generalized orthogonal designs~\cite{orth_designs} 
will be reported elsewhere.

\subsection{Perfect CSI at the Transmitter} \label{sec:perf_csi}
When perfect CSI is available at both the ends, the system equation can be written as 
\begin{eqnarray}
\label{perfect_ch_model}
\bmY = \sqrt{\frac{\rho}{N_t}} \hspp 
\sum_{k=1}^K  \bH \hsppp \bA_k \hsppp \bmx_k + \bmW 
\end{eqnarray} 
with a power constraint $\sum_{k=1}^K \trace \big( \bA_k \bA_k^{\dg}
\big) \leq N_t N_c$. 
Following~\cite{yibo_jiang,clin_ldcorr}, it can be shown that we have the following 
upper bound for the mutual information: 
\begin{eqnarray}
I( {\bmX}; {\bmY}| {\bmH} = \sfh) =  I(\bmx_1,\ldots,\bmx_K ; {\bmY}| {\bmH}= \sfh) 
\leq \sum_{k=1}^K \underbrace { I \left(\bmx_k; \sqrt{\frac{\rho}{N_t}} \hsppp 
{\bmH} \hsppp {\bA}_k \hsppp \bmx_k+ {\bmW} \left| {\bmH}= \sfh \right. 
\right). } _{I_k } 
\label{eq_constr1}
\end{eqnarray} 
Equality in (\ref{eq_constr1}) holds if and only if the GOC is satisfied. 
From~\cite{yibo_jiang,clin_ldcorr}, we also have 
\begin{eqnarray} 
\label{eqn_ik}
I_k & = & \varphi \left(\frac{\rho}{N_t} 
\trace \left(  \sfh \bQ_k \sfh^{\dg} \right) \right) - h( \bmn )
\end{eqnarray} 
where $\bQ_k = \bA_k \bA_k^{\dg}$, $\varphi(a) = h(\sqrt{a} \bmx + {\bmn} | 
{\bmH} = \sfh)$, $h(\cdot)$ denotes the differential entropy, 
and ${\bmn}$ is a real zero-mean Gaussian 
of variance $1/2$. The 
structure of the optimal LD code is as follows. 
\begin{thm} 
\label{prop_part1} 
Let $\bmX$ be an LD code as in~(\ref{LD_code}) with $K$ symbols and let the 
corresponding dispersion matrices be $\{ {\bA}_k \hsppp , \hsppp k = 1, 
\cdots , K \}$. Also, let the input symbols $\bmx_1, \cdots, \bmx_K$ satisfy 
Assumption~1. The instantaneous mutual information can be upper bounded as 
\begin{eqnarray}
I( {\bmX}; {\bmY}| {\bmH} = \sfh) \leq K \left[\varphi \left( 
\frac{\rho N_c}{K} \lambda_{\max}(\sfh^{\dg} \sfh) \right) - h(\bmn) \right] 
\end{eqnarray}
with equality if and only if $\{ \bA_k \}$ satisfy the GOC and $\bA_k \bA_k^{\dg} = 
\bQ_k = \bQ$ for all $k$ where $\bQ = \bU _{\sfh} \bfLambda_{\sfh} \bU_{\sfh}^{\dg}$. 
The matrix $\bU_{\sfh}$ is an eigenvector matrix of $\sfh^{\dg} \sfh$ (in the standard order) 
and the only non-zero entry in $\bfLambda_{\sfh}$ is the leading diagonal element 
whose value is $\frac{N_t N_c}{K}$. 
\end{thm} 
{\vspace{0.03in}}
\begin{proof} 
See Appendix~\ref{app_1}. 
\end{proof}

The above result shows that the optimal choice of $\{ \bQ_k \}$ is independent of 
$k$, that is, uniform symbol power allocation is {\em still} optimal. Furthermore, 
this scheme excites only the dominant spatial mode. 
A generic singular value decomposition for $\bA_k$ shows that it
has to satisfy $\bA_k = \sqrt{ \frac{N_t N_c}{K} } \hsppp \bU_{\sfh}[1] \bv_k$ where 
$\bU_{\sfh}[1]$ is $N_t \times 1$ and is the first column of $\bU_{\sfh}$, and $\bv_k$ 
is an $1 \times N_c$ vector of unit norm. With this structure, it can also be checked 
that the GOC can be met if and only if the $K \times N_c$ matrix $\bV$ defined as 
\begin{eqnarray}
\bV = \left[ \begin{array}{cccc} 
\bv_1^T & \bv_2^T &  \cdots & \bv_K^T \end{array} 
\right]^T \nonumber 
\end{eqnarray} 
satisfies $\bV \bV^{\dg} = \bI_K + i {\bf X}$ where ${\bf X}$ is real 
skew-symmetric\footnote{An $n \times n$ matrix ${\bf X}$ is said to be real 
skew-symmetric if it has real entries and satisfies ${\bf X}^T = -{\bf X}$.}. 
Based on this 
decomposition, we can completely characterize the structure of the dispersion 
matrices $\{\bA_k \}$ and can establish the existence of an LD code 
if and only if $K \leq 2 N_c$. The following proposition states this result. 

\begin{prop}
\label{prop_part2}
There exists a $K \times N_c$ matrix $\bV$ such that $\bV \bV^{\dg} = \bI_K + i {\bf X}$ 
where ${\bf X}$ is real skew-symmetric if and only if $K \leq 2 N_c$.
\endproof 
\end{prop} 

Due to space constraints, the proof of the claim and the explicit construction of 
the dispersion matrices are not reported here. 
However, we provide simple illustrations of these constructions now. 
When $K \leq N_c$, any set of $K$ rows of an arbitrary $N_c \times N_c$ 
unitary matrix works for $\bV$. Since $\{ \bv_k \}$ are pairwise orthogonal, 
the GOC is naturally met. In fact, it is to be noted that a stronger condition 
(than the GOC) holds: $\bA_k \bA_j^{\dg} = {\bf 0}$ for all $k \neq j$. Under 
the same conditions as above,~\cite{orth_designs} proposes further explicit 
constructions 
to meet the GOC. However, these conditions are only sufficient, but not necessary 
as the statement of Prop.~\ref{prop_part2} illustrates. 

The surprising\footnote{In retrospect, this claim is not all that surprising 
since the use of real input symbols means that $R$ bits can be transmitted per 
complex-dimension if a $2^R$-ary constellation is used for signaling.} 
claim of Prop.~\ref{prop_part2} is that the GOC can be met and the 
data-streams separated temporally as long as $K \leq 2 N_c$. The two-fold gain in 
the maximum possible choice of $K$ (which is $2 N_c$) over the `naturally' 
expected limit of $K = N_c$ stems primarily from the weaker condition that 
$2 \hsppp {\rm Re}(\bA_k \bA_j^{\dg}) = \bA_k \bA_j^{\dg} + \bA_j \bA_k^{\dg} = 
{\bf 0}$ for all $k \neq j$, instead of the more stringent condition that 
$\bA_k \bA_j^{\dg} = {\bf 0}$ for all $k \neq j$. For example, when $K = 2N_c$, the 
condition 
in Prop.~\ref{prop_part2} can be satisfied by choosing 
\begin{eqnarray}
{\bf v}_k = \left \{ 
\begin{array}{c}
\Big[ \begin{array}{ccccccc}
0 & \cdots & 0 & \underbrace{1}_{k_1} & 0 & \cdots & 0 
\end{array} \Big] \hsp {\rm if} \hspp {k} \hspp {\rm is} \hspp {\rm odd}, \hspp 
k_1 = \frac{k+1}{2} \\ 
\Big[ \begin{array}{cccccccc}
0 & \cdots & 0 & \underbrace{i}_{k_2} & 0 & \cdots & 0 
\end{array} \Big] \hsp {\rm if} \hspp {k} \hspp {\rm is} \hspp {\rm even}, \hspp 
k_2 = \frac{k}{2}
\end{array}
\right. 
\label{vproper_choice}
\end{eqnarray}
\ignore{ 
\begin{eqnarray} 
\label{vproper_choice}
\bV = \left[ \begin{array}{ccccc} 
1 & 0  & 0 & \cdots & 0 \\ 
i & 0 & 0 & \cdots & 0 \\ 
0 & 1 & 0 & \cdots & 0 \\ 
0 & i & 0 & \cdots & 0 \\ 
\vdots & \vdots & \vdots & \ddots & \vdots \\ 
0 & 0 & 0 & \cdots & 1 \\ 
0 & 0 & 0 & \cdots & i 
\end{array} \right].
\end{eqnarray}
}

We are now prepared to state the main result of this section. 
\begin{thm} 
\label{prop_mutual_increase_k} 
Consider the family of orthogonal LD codes whose decoding complexity is comparable 
with OSTBC. The mutual information achievable with such codes is a non-decreasing 
function of $K$ which implies that $K = 2 N_c$ is necessary for optimal signaling. 
Furthermore, the optimal signaling scheme reduces to beamforming along $\bU_{\sfh}[1]$. 
{\em More simply stated, not coding across time is optimal from an information theoretic 
perspective.} 
\end{thm} 
\begin{proof} 
First, note that the GOC has to be met for the case of orthogonal LD codes. 
From Prop.~\ref{prop_part2}, we see that $K = 2N_c$ is the largest value of 
$K$ such that this is possible. With $K = 2 N_c$ and ${\bf v}_k$ 
as in~(\ref{vproper_choice}), 
the average mutual information can be expressed as 
\beqa 
\nonumber I(\bmx_1,\ldots,\bmx_K ; {\bmY}| {\bmH}) 
& = & K \varphi \left( \frac{Z}{K}  \right) - Kh( \bmn ) 
\nonumber
\eeqa 
where $Z = \rho N_c  \lambda_{\max}(\sfh^{\dg} \sfh)$. Letting $K$ to be a 
continuous parameter in the previous expression, we observe that the 
derivative of the mutual information with respect to $K$ is positive. For 
this, note that for any $Z \geq 0$, we have 
\beqa 
\label{user}
\varphi \left( \frac{Z}{K} \right) - h(\bmn) = \int_{0}^{ \frac{Z}{K} } 
\varphi'(y) \hsppp dy \geq \frac{Z}{K} \hsppp \varphi ' \left( \frac{Z}{K}  \right) 
\eeqa
since $\varphi(\cdot)$ is a differentiable function with $\varphi(0) - h(\bmn) = 0$ 
and $\frac{d \varphi(a)}{da} = \frac{1}{2} \hsppp {\sf mse}(a)$ and hence, 
monotonically decreasing in $a$; see App.~\ref{app_1} for details. 
In this setting, 
we have 
\begin{eqnarray} 
\bY & = & \sqrt{ \frac{\rho N_c}{K} } \hsppp \bH \hsppp 
\bU_{\sfh}[1] \hsppp \bmx + \bmW \hspp 
= \hspp \bH \hsppp \bU_{\sfh}[1] \hsppp {\bf x}_{\sf trans}  + \bmW, \nonumber \\ 
\bmx_{\sf trans} & = & \sqrt{ \frac{\rho}{2} } \hsppp {\bf x}, \hsp 
{\bf x} = \left[ \bmx_1 + i \bmx_2 \hspp , \hspp 
\bmx_3 + i \bmx_4 \hspp , \hspp \cdots \hspp , \hspp 
\bmx_{2N_c-1} + i \bmx_{2N_c}    \right] \label{xtrans} 
\end{eqnarray} 
for the system equation. 
In other words, the optimal signaling scheme reduces to {\em beamforming the 
complex symbol $\left(\bmx_{2k-1} + i \bmx_{2k}\right)/\sqrt{2}$ along the fixed 
transmit direction $\bU_{\sfh}[1]$ in the $k$-th symbol period of the coherence block 
with a transmit energy of $\rho$.} The proof is complete. 
\end{proof}

\section{Quantized CSI at the Transmitter}
\label{sec:quan_csi} 
From the previous section, we see that rank-one signaling (beamforming) is optimal in 
the perfect CSI case while in the statistical case, the rank of the optimal scheme could 
be greater than one, in general. It is natural to expect a smooth transition 
in the rank as the quality of CSI gets refined with increasing $B$. In this section, 
we show that a rank-one scheme has strong optimality properties. 
This observation is based on the following two results: 1) When $B$ is sufficiently 
large (to be characterized more precisely soon), a rank-one scheme maximizes the 
average mutual information, 2) In the small $B$ regime, we can show that a 
rank-one scheme maximizes the average received $\snr$. 

We first make precise the notion of a $B$-bit limited feedback scheme in the context 
of LD codes. We assume the knowledge of a codebook (of $2^B$ codewords) at the 
transmitter and the receiver where each codeword is a set of $K$ dispersion matrices 
satisfying a total power constraint. That is, the codebook $\cb$ is 
\beq 
\label{eq_cbt}
\cb = \left\{ C^{\ell} = (\bA_1^\ell, \cdots, \bA_K^{\ell}) : 
\sum_{k=1}^K \trace( \bA_k^\ell {\bA_k^\ell}^\dg) \leq N_t N_c, \hspp 
\ell = 1, \cdots, 2^B  \right\}. 
\eeq 
If the $\ell$-th codeword is used in signaling, the system model is described by 
\beq
\label{quan_ch_model}
\bmY = \sqrt{\frac{\rho}{N_t}} \hspp 
\sum_{k=1}^K  \bH \hsppp \bA_k^{\ell} 
\hsppp \bmx_k + \bmW. 
\eeq 
Recall from the previous section that 
\beqa \nonumber
I(\bmx_1,\ldots,\bmx_K ; {\bmY}| {\bmH}= \sfh) 
& \leq &  \sum_{k=1}^K  I \left(\bmx_k; \sqrt{\frac{\rho}{N_t}} \hsppp 
{\bmH} \hsppp {\bA}_k^{\ell} 
\hsppp \bmx_k+ {\bmW} \left| {\bmH}= \sfh \right. 
\right) \\ 
& = & \sum_{k=1}^K \varphi \left(\frac{\rho}{N_t} 
\trace \left(  \sfh \bQ_k^{\ell} \sfh^{\dg} \right) \right) - Kh( \bmn )
\label{eq_upperbound_parcsi}
\eeqa 
where the upper bound is met if $\{ \bA_k^{\ell} \}$ satisfies the GOC, 
$\bQ_k^{\ell} = \bA_k^{\ell} {\bA_k}^{\ell\dg}$ 
and $h(\bmn)$ is defined as in (\ref{eqn_ik}). 
Thus, the mutual information is completely characterized by the set of covariance 
matrices $(\bQ_1^{\ell}, \cdots , \bQ_K^{\ell})$. 
Over each coherence block, the receiver feeds back $\ell^{\star}$, the index of the 
optimal codeword which maximizes the instantaneous mutual information to the 
transmitter, and the transmitter communicates over the remaining symbols in the 
coherence block according to~(\ref{LD_code}) with dispersion matrices 
$\{ \bA_k^{\ell^{\star}} \}$. 
We now show that uniform symbol power allocation is optimal even in the partial CSI 
case. 
\begin{prop} 
\label{thm_parcsi_time_opt} 
Let the $B$-bit quantized feedback system be described as in~(\ref{quan_ch_model}). 
For any choice of $B$, the average mutual information is maximized by a 
codebook that allocates uniform power to input symbols. In fact, for any codeword 
index $\ell$, we have $\bQ_k^\ell = \bQ^\ell$ for all $k$. 
\end{prop}
\begin{proof} 
See Appendix~\ref{app_3}. 
\end{proof}

Thus, from the above theorem, we only need to quantize $\bQ^{\ell}$. The most 
natural quantization for a covariance matrix is based on an eigen-decomposition 
of $\bQ^{\ell}$. For this, we let $N_1$ and $N_2$ be such that $N_1 N_2 = 2^B$, 
and quantize $\bQ^{\ell}$ as 
\beq
\label{eq_quan_q}
\bQ^{\ell} = \bQ^{(i-1)N_2+j} = \bU_i \bfLambda_j \bU_i^\dg  
\triangleq \bQ^{i,\hsppp j} , \hsp i = 1, \cdots, N_1 \hspp {\rm and} \hspp 
j = 1, \cdots, N_2 
\eeq
where $\{ \bU_i \}$ are unitary and $\{ \bfLambda_j \}$ are positive semi-definite 
diagonal with $\trace (\bfLambda_j) \leq N_t N_c/K$. 
While the above quantization seems natural, it is unclear how to allocate $B$ into 
$N_1$ and $N_2$ {\em optimally}. In the ensuing discussion, we consider two cases 
and study the optimality of rank-one codebooks. 

\subsection{Strong Optimality of Rank-One Codebooks when $N_2 \geq N_t$}
Since the rank of $\bQ^{i, \hsppp j}$ is 
that of $\bfLambda_j$, 
we fix $\{ \bU_i \}_{i=1, \hsppp \cdots , \hsppp N_1}$ to be a known family 
and optimize over all possible 
$\{\bfLambda_j\}.$ This results in the following optimization: 
\begin{eqnarray} 
\label{eq_optproblem_lambda}
\bfLambda_{j^{\star}} = \max_{ \left \{\bfLambda_j \hsppp : \hsppp 
\trace( \bfLambda_j ) \hsppp \leq \hsppp \frac{N_t N_c}{K} \right \} }
\bEe \left[ \max_{i,j} \varphi \left( \frac{\rho}{N_t} 
\trace (\sfh \bU_i \bfLambda_j \bU_i^\dg \sfh^\dg) \right) \right].
\end{eqnarray} 
We say that rank-one codebooks are {\em strongly optimal} if 
a codebook of rank-one power allocations is sufficient to maximize the average 
mutual information for any choice of $\{ \bU_i \}$. 

\begin{thm} 
\label{thm_parcsi_opt_rk1_lg}
The main conclusion is that if $N_2 \geq N_t$, a rank-one codebook is {\em strongly 
optimal}. 
\end{thm}
\begin{proof} 
For each realization of the channel $\bmH = \sfh$, we seek to maximize the 
instantaneous mutual information by choosing the optimal codeword at the 
receiver and feed back the index $(i^{\star}, \hsppp j^{\star})$ to the transmitter 
through the feedback link. For this, note that for any fixed $\{\bU_i\}$,
\begin{eqnarray}
\max_{i,j} \varphi \left( \frac{\rho}{N_t} \trace (\sfh \bQ^{i,\hsppp j} \sfh^\dg) \right) 
& \stackrel{(a)}{=} & \max_{i,j} \varphi \left( \frac{\rho}{N_t} 
\trace (\bS_i \bfLambda_j \bS_i^\dg ) \right) 
\hsp \stackrel{(b)}{=}  \max_{i,j} \varphi \left( \frac{\rho}{N_t} 
\sum_{m=1}^{N_t} \bfLambda_j(m) \| \bs_{i m} \|^2   \right) \nonumber \\
& \stackrel{(c)}{=} &  \max_{i,j} \varphi \left( \frac{\rho N_c}{K} 
\sum_{m=1}^{N_t} \alpha_{jm} s_{im}   \right) 
\stackrel{(d)}{\leq}  \varphi \left( \frac{\rho N_c}{K} s_{i^{\star} m^{\star}} 
\right) \nonumber 
\end{eqnarray}
where (a) follows by defining $\bS_i = \bfLambda_\sfh^{1/2} \bU_\sfh^\dg \bU_i$ and 
$\sfh^\dg\sfh = \bU_{\sfh} \bfLambda_{\sfh} \bU_{\sfh}^\dg$, (b) and (c) 
follow by denoting the $m$-th column of $\bS_i$ by $\bs_{i m}$, its norm 
by $s_{im}$, $\alpha_{jm} = \frac{ \bfLambda_j(m) \hsppp K }{N_t N_c}$ and 
$\sum_m\alpha_{jm} \leq 1$, and (d) follows 
by letting $(i^{\star}, m^{\star}) =  \arg \max_{1 \hsppp \leq \hsppp i \hsppp \leq \hsppp N_1, 
\hspp 1 \hsppp \leq \hsppp m \hsppp \leq \hsppp N_t} s_{im}$. 
If $N_2 \geq N_t$, we can consider a distinct set of $N_t$ rank-one power allocations 
each of which excites only one mode. Using this set in the above framework allows us 
to meet the upper bound. 
\end{proof} 

Note that the condition $N_2 \geq N_t$ implies that $2^B = N_1 N_2 \geq N_t$. But 
this inequality does not impose any constraint on $N_1$. Nevertheless, we can 
say that if $B$ is sufficiently large ($B \gg \log(N_t)$) so that at least 
$\log(N_t)$ bits can be allocated to quantize the power allocation component of 
$\bQ^{\ell}$, rank-one codebooks are always optimal irrespective of the 
constellation of input symbols, $\snr$, channel correlation etc. This optimality 
is not completely surprising since the quantized feedback system 
closely approximates a perfect feedback system when $B \gg \log(N_t)$. 

\subsection{Weak Optimality of Rank-One Codebooks when $N_2 < N_t$}
From the notation of Theorem~\ref{thm_parcsi_opt_rk1_lg}, when $N_2 < N_t$ 
we can rewrite the optimization in~(\ref{eq_optproblem_lambda}) as
\beq
\label{eq_optproblem_lambda_modify}
\max_{ \left\{\alpha_{jm} \hsppp  : \hsppp \sum_{m=1}^{N_t} \alpha_{jm} \hsppp 
\leq \hsppp 1 \hsppp {\rm for} \hsppp {\rm all} \hsppp j \right\} } 
\bEe \left[ \max_{i,j} \varphi \left( \frac{\rho N_c}{K} 
\sum_{m=1}^{N_t} \alpha_{jm} s_{im}   \right) \right].
\vspp
\eeq 
Direct optimization of (\ref{eq_optproblem_lambda_modify}) requires the exact 
distribution function of $s_{im}$ which is a complicated function of the 
spatial correlation, thus rendering the above problem intractable. We now consider 
an alternate formulation of the above problem wherein the objective function is 
the minimization of $\Delta I$ with 
\begin{eqnarray} 
\Delta I & = & 
\bEe \left[ \varphi \left( \frac{\rho N_c}{K} \lambda_{\max}(\sfh^{\dg} \sfh) \right)
 - \max_{i,j} \varphi \left( \frac{\rho N_c}{K} 
\sum_{m=1}^{N_t} \alpha_{jm} s_{im}   \right)
\right]. 
\nonumber 
\end{eqnarray}
That is, the objective is to minimize the difference in average mutual information 
between the perfect CSI benchmark and a quantized feedback scheme. We now propose an 
upper bound for $\Delta I$ that renders the study of optimal signaling tractable 
in a weak sense. 
\begin{lem}
\label{lem_mse}
The quantity $\Delta I$ can be upper bounded by $\Delta \snr$, the difference 
in average received $\snr$, defined as 
$\Delta \snr \triangleq \frac{\rho N_c}{K} \cdot 
\bEe_{\sfh} \left[ \lambda_{\max}(\sfh^H \sfh) - \sum_{m=1}^{N_t} \alpha_{jm} s_{im} 
\right].$
\end{lem}
\begin{proof} 
See Appendix~\ref{app_lem_mse}. 
\end{proof}
Thus, in a weak sense, the optimization in~(\ref{eq_optproblem_lambda_modify}) is 
equivalent to maximizing the average received $\snr$ of the quantized feedback 
scheme: 
\beq
\label{eq_optproblem_lambda_modify_upper}
\max_{ \left\{\alpha_{jm} \hsppp  : \hsppp \sum_{m=1}^{N_t} 
\alpha_{jm} \hsppp \leq \hsppp 1  \hsppp {\rm for} \hsppp 
{\rm all} \hsppp j \right\} } \bEe \left[  \frac{\rho N_c}{K} \max_{i,j}
\sum_{m=1}^{N_t} \alpha_{jm} s_{im}   \right].
\eeq
With this new metric, we now establish the optimality of rank-one codebooks. 
\begin{thm} 
\label{thm_onemode_quan}
Let the $B$-bit quantized feedback system and the corresponding $N_1$ and $N_2$ 
be described as before. If $N_2 < N_t$, the average $\snr$ at the receiver is 
maximized by a rank-one codebook.
\end{thm}
\begin{proof}
See Appendix~\ref{app_onemode_quan}. 
\end{proof}
It is important to note that the optimality of a rank-one codebook in terms of the 
new metric {\em does not} necessarily imply the optimality of rank-one codebooks in 
terms of the average mutual information. For any choice of correlation and $B$ that 
is comparable to $\log(N_t)$, we expect a natural transition from strong optimality 
to weak optimality as $\snr$ increases. Nevertheless, numerical studies suggest that 
this transition $\snr$ is very large for most reasonable correlation, so that practically 
speaking {\em rank-one codebooks are still optimal.} This will be the focus of our 
future work. 
Furthermore, following the approach in Theorem~\ref{prop_mutual_increase_k}, 
the system equation reduces to 
\begin{eqnarray}
\label{heree}
\bY & = &  \bH \hsppp \bU_{i^{\star}}[j^{\star}] \bmx_{\sf trans} + \bmW 
\end{eqnarray} 
where ${\bf x}_{\sf trans}$ is as in~(\ref{xtrans}) and for any given realization 
$\bH = \sfh$, $\bQ^{i^{\star}, j^{\star}} = \bU_{i^{\star}} \bfLambda_{j^{\star}} 
\bU_{i^{\star}}^\dg$ with $\bfLambda_{j^{\star}} = \frac{N_t N_c}{K} \hsppp 
\diag (e_{j^{\star}})$ and $e_{j^{\star}}$ is the $j^{\star}$-th standard basis vector 
of ${\mathbb R}^{N_t}$. 

It is critical to note the difference between a beamforming 
scheme in the classical sense and the system equation in~(\ref{heree}). In the classical 
sense, the beamforming direction is {\em fixed and independent} of the channel 
state, but perhaps dependent on the channel statistics which evolves over slower 
time scales. In~(\ref{heree}), the 
beamforming direction is dependent on the channel state and is based on the feedback 
information. Despite the adaptation of this direction in response to the reverse 
link feedback, the low-complexity gain associated with beamforming (in the classical 
sense) can be accrued because we still need only a single radio link chain to implement 
this scheme. The need to adapt the beamforming direction at the transmitter at a 
fast rate\footnote{The rate has to be slightly faster than the rate at which the 
channel evolves.} may impose additional constraints on the hardware, but these are 
expected to 
be sub-dominant in comparison with the performance improvement obtained by utilizing 
the feedback information. 

To summarize, the main conclusion of this work is: {\em Coding across time is not 
necessary to maximize the average mutual information if the quality of CSI feedback 
is sufficiently good; The same conclusion holds with a low quality of CSI feedback 
if the objective is relaxed to that of maximizing the average received $\snr$.}

\section{Simulation Results} 
\label{sec_num} 
We now present numerical studies to demonstrate that the rank-one codebook is a 
reasonable choice for most practical scenarios of interest. 
We study three settings here: 1) a $2 \times 2$ i.i.d.\ channel, 2) a $4 \times 4$ 
i.i.d.\ channel, and 3) a $4 \times 4$ correlated channel with variance of channel 
entries given by 
\beq \nonumber
{\rm V}_4 = \frac{16}{2.6}\left[ \begin{array}{cccc}
0.1 & 0 & 0.4 & 0\\
0 & 0.1 & 0.4 & 0\\
0 & 0 & 0.4 & 0.4\\
0 & 0 & 0.4 & 0.4
\end{array} \right]. \nonumber 
\eeq 
{\vspace{0.05in}}
{\noindent}In all the cases, the channel power $\bEe[\trace(\bH \bH^{\dg})]$ 
is normalized as $N_t N_r$. 
For all cases, we compare the mutual information between the best rank-one 
and `best' rank-two codebooks. The cases studied are: a) $B = 2$, $N_1 = 4$, 
$N_2 = 1$, and b) $B = 2$, $N_1 = 2$, $N_2 = 2$. 
We now elaborate on how to obtain the best rank-one and rank-two codebooks. 

From~(\ref{eq_quan_q}), the structure for each codeword is 
$\bQ^{i,\hsppp j} = \bU_i \bfLambda_j \bU_i^\dg.$ 
Fixing $\{\bU_i\}$, the different rank-one codebooks are characterized by different 
choices of rank-one $\{\bfLambda_j\}$. For example, with $N_r = N_t = 4$, $N_1 = 4$ 
and $N_2 = 1$, there are four choices of rank-one codebooks: ${\bf \Lambda}^i = 
\diag( e_i )$ where $e_i$ is the $i$-th standard basis vector of ${\mathbb R}^4$. 
Similarly, there are six possible rank-one choices for $N_r = N_t = 4$, $N_1 = 2$ 
and $N_2 = 2$. The best rank-one codebook is the one that maximizes the average mutual 
information. 
The above procedure is difficult to extend for rank-two codebooks. This 
is because even though there are only finite choices for the positions of the modes 
that can be excited, the power allocations between these two excited modes can run 
through a continuum. For example, with $N_r = N_t = 4$, $N_1 = 4$ and $N_2 = 1$, 
$\bfLambda_1 = \diag(1/2,\hsppp 1/2, \hsppp 0,\hsppp 0)$, $\bfLambda_1 = 
\diag(1/3,\hsppp 2/3, \hsppp 0, \hsppp 0)$, or $\bfLambda_1 = \diag(0,\hsppp 
2/3, \hsppp 0, \hsppp 1/3)$ are all feasible choices for rank-two codebooks. 
This difficulty forces us to study this case by randomly generating $50$ different 
sets of $\{\bfLambda_j\}$ and picking the `best' rank-two codebook from this random set. 
Further, since there is no proper distance metric to pack unitary matrices, a 
random family of $\{ \bU_i \}$ are generated via random vector quantization (RVQ). 
Numerical studies show that there is roughly very 
similar performance with different choices of $\{ \bU_i \}$ 
and hence, only one such choice is highlighted.

In the simulations, the choice of $K$ used is $N_c$. This is because while rank-one 
codebooks meeting the GOC exist for up to $K = 2 N_c$, the study in Sec.~\ref{sec_stat} 
suggests that rank-two codebooks that meet the GOC may not exist. We illustrate our 
results with Gaussian inputs, but numerical studies show that input constellation 
plays a minimal role in the trends. 

\begin{figure}
\centering
\includegraphics[height=6in,width=6in]{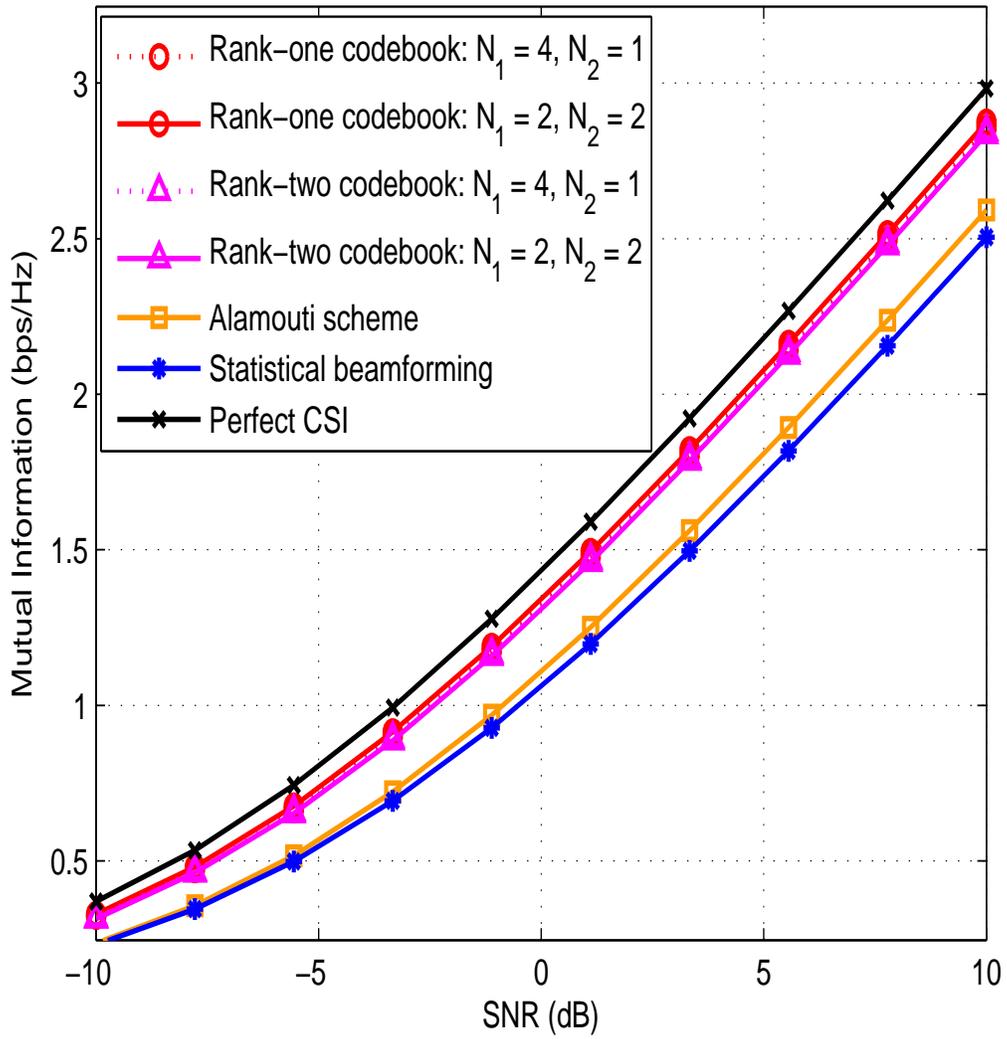}
\caption{Average mutual information of different codebooks in a $2 \times 2$ i.i.d.\ 
channel.}\label{iid_2by2_Gaussian}
\end{figure}

\begin{figure}
\centering
\includegraphics[height=6in,width=6in]{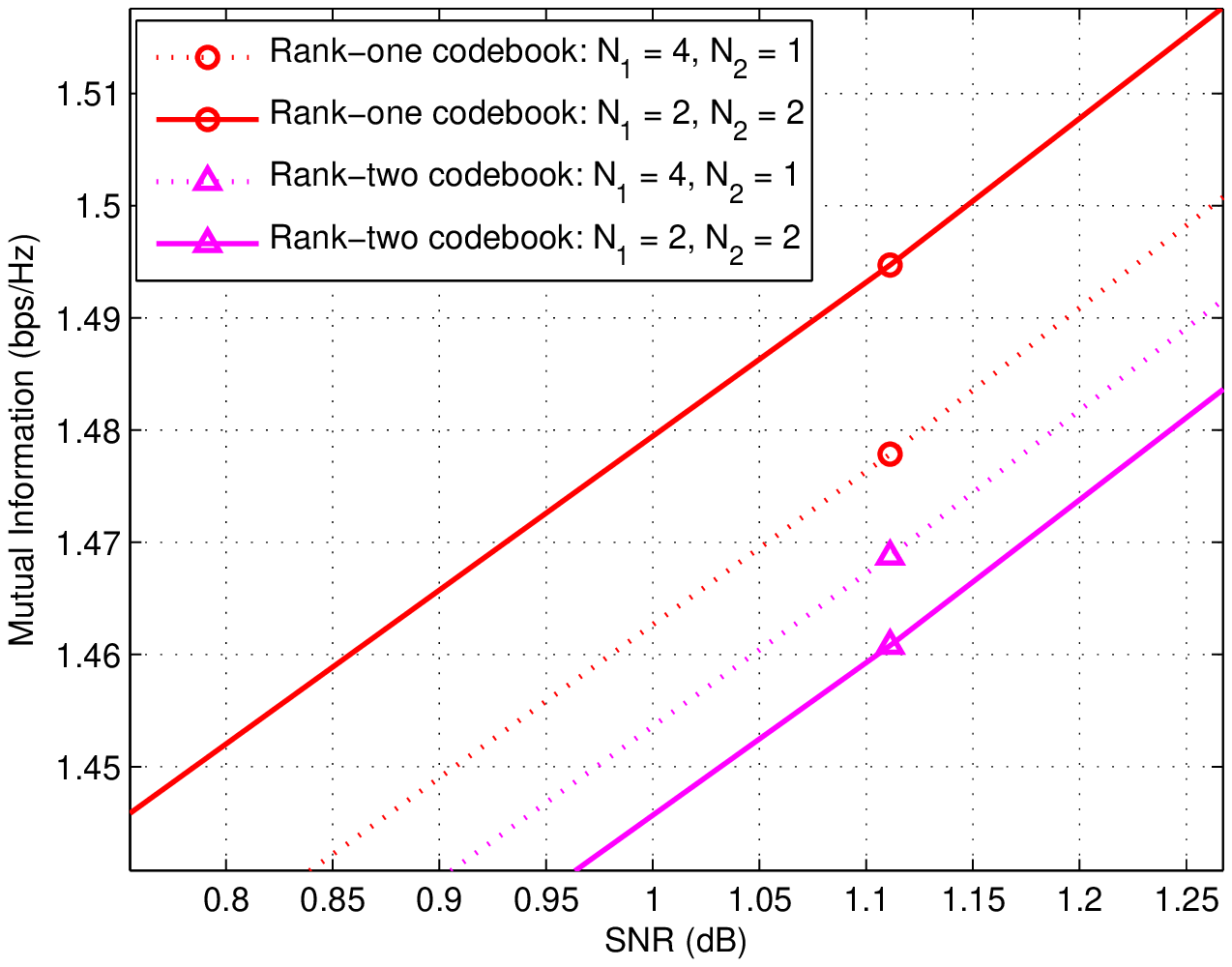}
\caption{A magnified view of the performance of rank-one and rank-two codebooks 
in the $2 \times 2$ i.i.d.\ case.} \label{iid_2by2_Gaussian_detail}
\end{figure}

\begin{figure}
\centering
\includegraphics[height=6in,width=6in]{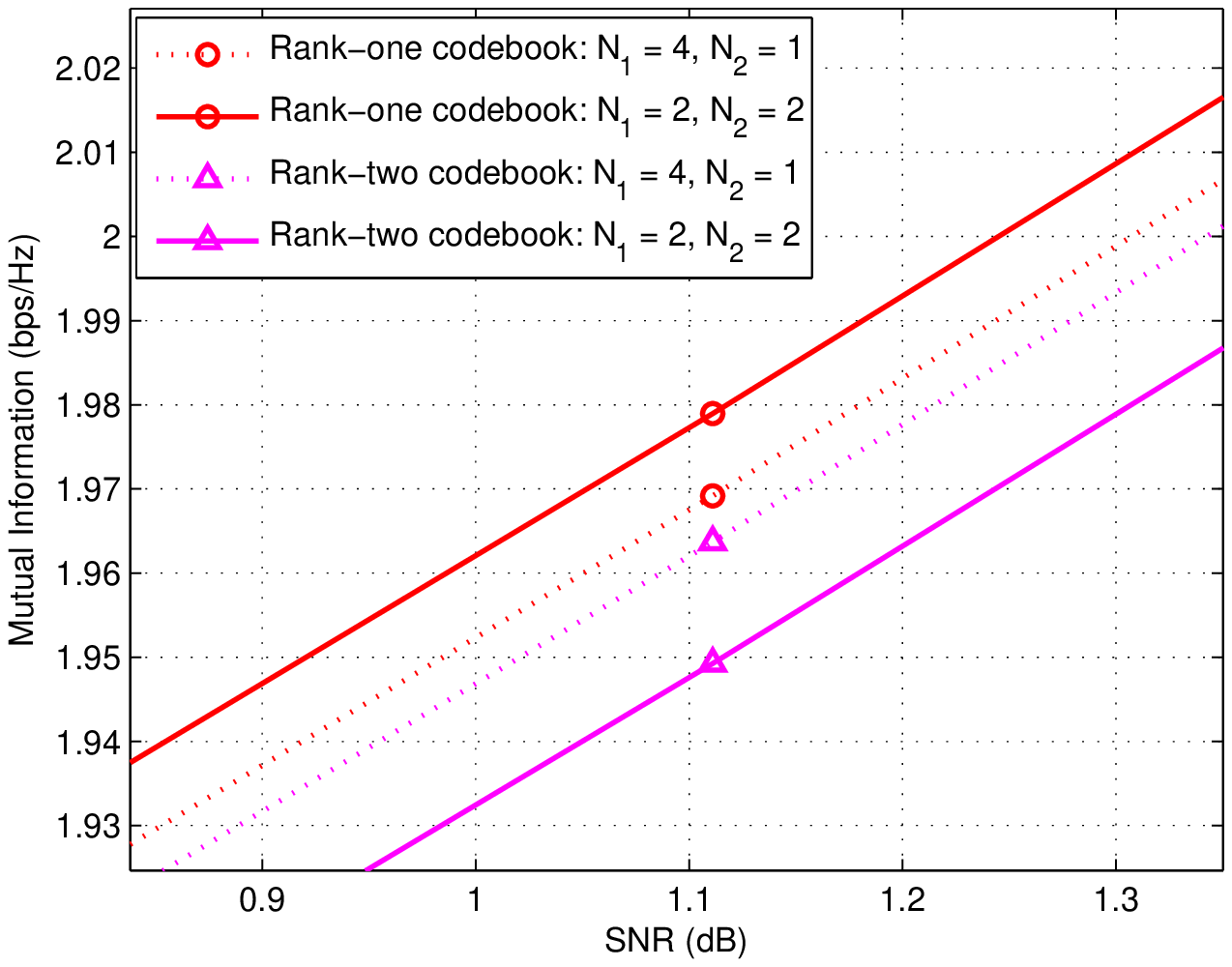}
\caption{A magnified view of the performance of rank-one and rank-two codebooks 
in the $4 \times 4$ i.i.d.\ case.} 
\label{iid_4by4_Gaussian_detail}
\end{figure}

\begin{figure}
\centering
\includegraphics[height=6in,width=6in]{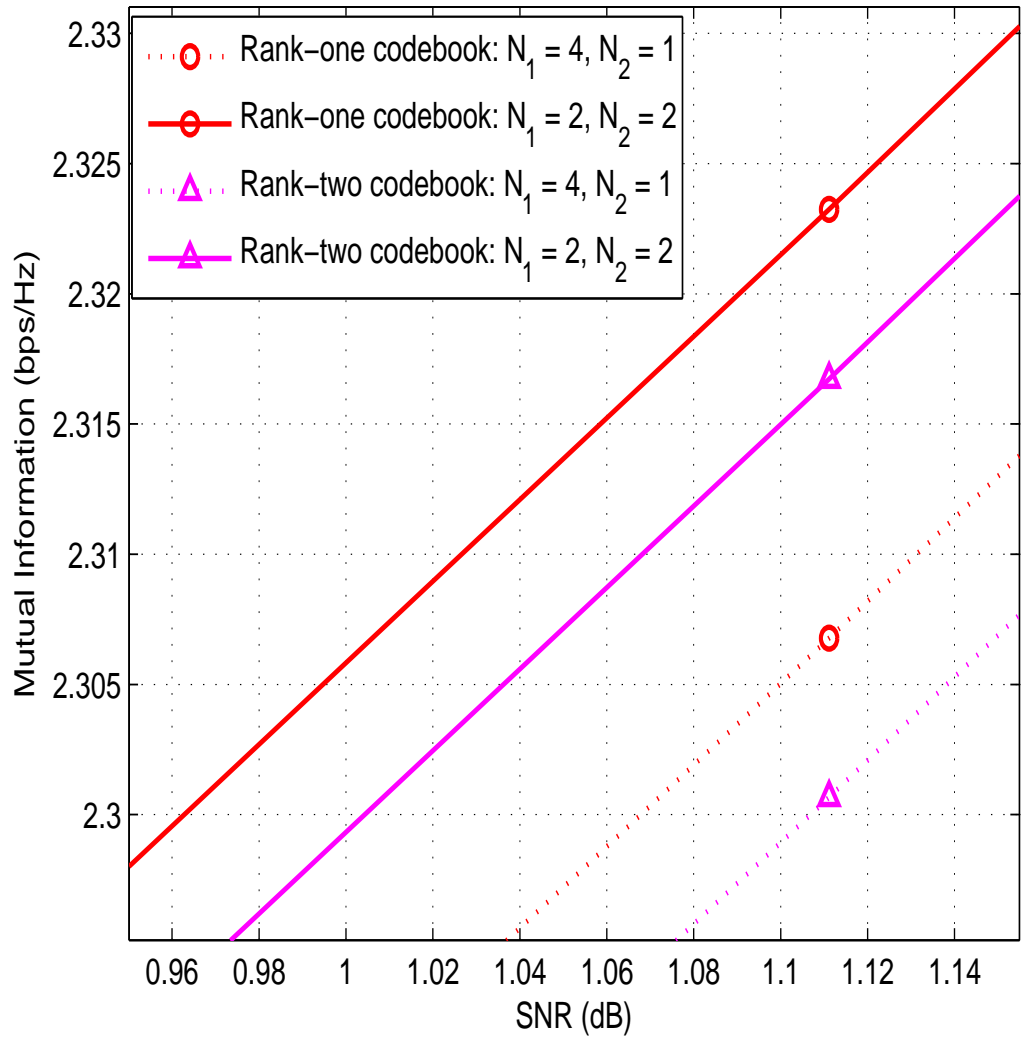}
\caption{A magnified view of the performance of rank-one and rank-two codebooks 
in the $4 \times 4$ correlated case.} 
\label{scatIII_4by4_Gaussian_detail}
\end{figure}

Fig.~\ref{iid_2by2_Gaussian} plots the mutual information with the best rank-one 
and rank-two codebooks for $N_1 = 4$, $N_2 = 1$, and for $N_1 = 2$, $N_2 = 2$. 
Benchmark plots of the perfect CSI (upper bound), only statistical information 
(lower bound), and statistical beamforming (lower bound) are also presented. A 
magnified view 
of this plot in Fig.~\ref{iid_2by2_Gaussian_detail} shows that the best rank-one 
codebook outperforms all other rank-two codebooks. In all subsequent plots, we focus 
only on a magnified view of the comparison between rank-one and rank-two codebooks 
since all plots show very similar trends for the mutual information and 
the main focus is on our conjecture that a rank-one scheme leads to good 
performance in practice. Fig.~\ref{iid_4by4_Gaussian_detail} and 
Fig.~\ref{scatIII_4by4_Gaussian_detail} both verify that a rank-one codebook outperforms 
rank-two codebooks in the $4 \times 4$ i.i.d.\ and $4 \times 4$ correlated channels, 
respectively, thus suggesting that in 
most practical scenarios of interest, beamforming is a good candidate for optimal 
signaling. 


\section{Conclusion}
In this work, we have studied the cases of coding across space and across 
space-time in a unified fashion by considering a family of linear dispersion 
codes that satisfy an orthogonality constraint. {\em Our results show that there is 
no need to code across time either when the channel information at the transmitter 
is perfect or when the channel information is of a sufficiently good quality.} 
On the other hand, even when the channel information is not of a good quality 
(corresponding to low rates of feedback), the low-complexity beamforming scheme 
possesses some attractive optimality properties, namely, it maximizes the 
average received $\snr$. From a design viewpoint, beamforming is particularly 
attractive: The low-complexity of its design augmented with the low-cost ensured by 
using a single radio link chain. 

Note that the orthogonal LD codes are of a complexity comparable to the OSTBC which 
are commonly used in standardization efforts. However, even in the 
case of rank-one signaling, one may be able to send $K > 2 N_c$ data-streams with 
dispersion matrices that do not meet the orthogonality constraint. The obvious 
disadvantage of this strategy is that the data-streams may have to be separated at 
the receiver with more complex decoding architectures. More so, the objective of 
maximizing the average mutual information of the inner space-time code can be 
met by {\em precoding schemes} that multiplex more than one data-stream, albeit at the cost 
of some decoding complexity. Thus, it is not clear as to what is the trade-off between 
mutual information and decoding complexity. Furthermore, our work provides a good 
justification as to why there has been significant recent attention on limited feedback 
precoding$/$beamforming 
schemes~\cite{david_grass,kiran_outage,vincent_lau,bhaskar_rao,david_mux,bhaskar_rao2,david_heath_multimode} 
rather than on limited feedback space-time coding schemes. 


Much work needs to be done to understand how these results translate 
to more practical scenarios of interest where the channel information at the receiver 
or the statistical information at the transmitter may not be 
perfect, the channel is not block fading, wideband etc. Construction of dispersion 
matrices that satisfy desired low-complexity properties is another area of interest. 
It is also important 
to note that we have only scratched the surface on understanding the trade-off between 
reliability and throughput with constraints on the complexity of the encoder-decoder 
pair. While reliability is an important design metric in certain situations, 
throughput (more coarsely identified as the `multiplexing gain') is probably a more 
important 
aspect in the design of high data-rate wireless systems. In such settings, it is of 
interest to understand how and when low-complexity, adaptive signaling techniques 
can be leveraged to achieve near-optimal performance.

\appendix 
\subsection{Proof of Theorem \ref{thm_stat_opt}} 
\label{app_2} 
Denote ${\widetilde{\bA}}_k {\widetilde{\bA}}_k^{\dg}$ by $\widetilde{\bQ}_k$ 
and $\frac{1}{K} \sum_{k=1}^K \widetilde{\bQ}_k$ by $\widehat{\bQ}$, and observe that 
$\widehat{\bQ}$ is a positive semi-definite matrix with trace constrained by 
$\frac{N_t N_c}{K}$. With $\gamma \triangleq \bEe \left[ I( {\bmX}; {\bmY}| 
{\bmH} = \sfh) \right] +K \hsppp \bEe \left[ h(\bmn)  \right]$, we have 
\begin{align}
\begin{split} 
\gamma & 
{\hspace{0.15in}} \stackrel{(a)}{\leq}  \bEe \left[ \sum_{k=1}^K I_k   \right] 
+ K \hsppp \bEe \left[ h(\bmn)  \right] = 
\bEe \left[ \sum_{k} \varphi \left( \frac{\rho}{N_t} \hsppp 
\trace \left( \widetilde{\bQ}_k \sfh^{\dg} \sfh \right) \right) \right]  
\nonumber \\ 
& {\hspace{0.15in}} 
\stackrel{(b)}{\leq}  K \cdot \bEe \left[ \varphi \left( \frac{\rho}{N_t K} 
\hsppp \sum_k \trace \left ( \widetilde{\bQ}_k \sfh^{\dg} \sfh \right) \right)
\right] 
= K \cdot \bEe \left[ \varphi \left( \frac{\rho}{N_t} 
\hsppp \trace \left ( \widehat{\bQ} \hsppp \sfh^{\dg} \sfh \right) \right)
\right] 
\end{split}
\end{align} 
where equality holds in (a) if the GOC condition is satisfied and (b) follows 
from the concavity of $\varphi(\cdot)$. Optimizing over the choice of 
$\widehat{\bQ}$ in the above equation 
results in an upper bound for 
$\bEe \left[ I( {\bmX}; {\bmY}| {\bmH} = \sfh) \right]$. Denote by $\bQ_{\opt}$ 
the solution to the following optimization problem: 
\begin{eqnarray} 
\label{eq_inst_opt_equi} 
\bQ_{\opt} = \arg\max _{ \trace(   \bQ)  \hsppp = \hsppp \frac{N_tN_c}{K}} \bEe 
\left[ \varphi \left(\frac{\rho}{N_t} 
\trace \left( \sfh \bQ \sfh^{\dg} \right) \right) \right] 
\end{eqnarray}
where $\bQ = {\widetilde{\bA}} {\widetilde{\bA}}^{\dg}$. A choice of dispersion 
matrices $\{ {\widetilde{\bA}}_k \}$ that satisfies ${\widetilde{\bA}}_k 
{\widetilde{\bA}}_k^{\dg} = \bQ_k = \bQ_{\opt}$ for all $k$ and that meets 
the GOC condition would result in an equality in the upper bound 
and hence achieves the ergodic capacity. The fact that $\bQ_{\opt}$ in
(\ref{eq_inst_opt_equi}) is diagonal follows from~\cite{clin_ldcorr}. 
\endproof

\subsection{Proof of Theorem~\ref{prop_part1}} 
\label{app_1} 
The connection between minimum mean squared error (MMSE) estimation and mutual 
information established in~\cite{guo_mmse} implies that 
$\frac{d\varphi(a)}{da} = \frac{1}{2} \hspp {\sf mse}(a)$ 
where ${\sf mse}(a)$ is the mean squared error for the channel under consideration 
at an $\snr$ of $a$. 
The positivity and the monotonous decrease of the ${\sf mse}(\cdot)$ function implies 
that $\varphi(\cdot)$ (and hence $I_k(\cdot)$) is concave and non-decreasing. 
We first upper bound $I_k$ and this leads to an upper bound on 
$I( {\bmX}; {\bmY}| {\bmH} = \sfh)$. For this, note that 
\begin{eqnarray}
I_k + h(\bmn)  
= \varphi \left(  \frac{\rho}{N_t} \trace \left(  \bQ_k \sfh^{\dg} \sfh \right) 
\right) 
= \varphi \left( \frac{\rho}{N_t} \sum_{i=1}^{N_t } 
\lambda_i (\bQ_k \sfh^{\dg} \sfh) \right) 
\stackrel{(a)}{\leq}
\varphi \left(  \frac{\rho}{N_t}\lambda_1(\sfh^{\dg} \sfh) \trace(\bQ_k) \right) 
\end{eqnarray}
where (a) follows from the fact that $\lambda_i( \bQ_k \sfh^{\dg} \sfh  ) \leq 
\lambda_i(\bQ_k) \lambda_1(\sfh^{\dg} \sfh)$ and the monotonicity of $\varphi(\cdot)$. 
The concavity of $\varphi$ implies that 
$\eta = \sum_{k=1}^K I_k  + K h(\bmn)$ satisfies 
\begin{eqnarray}
\eta \leq \sum_{k = 1}^K 
\varphi \left( \frac{\rho}{N_t}\lambda_1(\sfh^{\dg} \sfh)  \trace(\bQ_k) \right)
\leq K \hsppp  \varphi \left(  \frac{\rho}{N_t} 
\hsppp \frac{N_t N_c}{K} \hsppp \lambda_1(\sfh^{\dg} \sfh)  \right) 
= \varphi \left( \frac{\rho N_c}{K} \hsppp \lambda_1(\sfh^{\dg} \sfh) \right). 
\label{eqn_major}
\end{eqnarray} 

We now show that the upper bound is in fact achievable. Consider the maximization of 
$\sum_{k}I_k$ over the set ${\cal Q} =  \left \{ \bQ_k = 
\bQ \hspp {\rm for \hspp all} \hspp k, \hspp \bQ \succ \bo \hspp {\rm and} 
\hspp \trace(\bQ) = \frac{N_t N_c}{K} \right \}$. We then have 
$\omega = \varphi \left( \frac{\rho}{N_t} \trace \left(  \sfh \bQ \sfh^{\dg} \right) 
\right)$ satisfying 
\begin{eqnarray}
\omega & = & 
\varphi \left(  \frac{\rho}{N_t} \trace \left(  \bQ \sfh^{\dg} \sfh \right) \right) 
= \varphi \left( \frac{\rho}{N_t} \sum_{i=1}^{N_t } \lambda_i (\bQ \sfh^{\dg} \sfh)
\right) 
\stackrel{(a)}{\leq}  \varphi \left( \frac{\rho}{N_t} 
\sum_{i=1}^{N_t} \lambda_i(\bQ) \lambda_{i}(\sfh^{\dg} \sfh) \right)
\nonumber \\ & \stackrel{(b)}{\leq} & 
\varphi \left (\frac{\rho}{N_t}\cdot \frac{N_t N_c} {K} 
\hsppp \lambda_1(\sfh^{\dg} \sfh) \right) 
= \varphi \left( \frac{\rho N_c}{K} \hsppp \lambda_1( \sfh^{\dg} \sfh) \right) \nonumber 
\end{eqnarray}
where (a) follows from the monotonicity of $\varphi(\cdot)$ and the 
fact 
that if ${\bf A}$ and ${\bf B}$ are $n \times n$ positive semi-definite matrices, 
then $\sum_{i=1}^n \lambda_i({\bf A B}) \leq \sum_{i=1}^n \lambda_i({\bf A}) 
\lambda_i({\bf B})$ 
and (b) follows from trivially upper bounding $\lambda_i(\sfh^{\dg} \sfh)$ with 
$\lambda_1(\sfh^{\dg} \sfh)$. Also note that the upper bound is achieved 
by beamforming along the dominant eigen-direction of $\sfh^{\dg} \sfh$. 
Thus, we have 
\begin{eqnarray}
I( {\bmX}; {\bmY}| {\bmH} = \sfh) \leq \sum_{k = 1}^K I_k 
= K \varphi \left( \frac{\rho N_c}{K} \hsppp \lambda_1( \sfh^{\dg} \sfh) \right) 
- K h( {\bf n}) 
\end{eqnarray}
with equality if and only if $\bQ$ is as above and the GOC conditions are satisfied. 
\endproof

\subsection{Proof of Theorem~\ref{thm_parcsi_time_opt}} 
\label{app_3}
First, note that any (generic) codebook can be written as 
\beq
\label{eq_cb}
C = \left\{ c^\ell,  \hspp \ell = 1, \cdots, 2^B  \right\} 
{\hspp} {\rm where} \hspp 
c^{\ell} \triangleq (\bQ_1^\ell, \cdots, \bQ_K^{\ell}) \hspp {\rm s.t.} \hspp 
\sum_{k=1}^K \trace( \bQ_k^\ell) \leq N_t N_c. \nonumber
\eeq
Further, define a codebook $D$ as 
\beq
\label{eq_cb}
D = \left\{ d^\ell, \hspp \ell = 1, \cdots, 2^B  \right\} 
\hspp {\rm where} \hspp d^{\ell} \triangleq 
(\bQ^\ell, \cdots, \bQ^{\ell}) \hspp {\rm s.t.} \hspp \trace( \bQ^\ell) \leq 
\frac{N_t N_c}{K}. \nonumber 
\eeq 
Denoting the families of codebooks of the type $C$ and $D$ by ${\cal C}$ and 
${\cal D}$, respectively, we have ${\cal D} \subset {\cal C}$. 
With a codebook $C$ from ${\cal C}$, 
the average mutual information is 
\begin{eqnarray}
 \bEe_{ \sfh} \left[  \max_{ i}  
\sum_{k=1}^K \varphi \left( \frac{\rho}{N_t} \trace(\sfh \bQ_k^i \sfh^\dg) 
\right) \right] 
& \stackrel{(a)}{\leq} &  
\bEe_{\sfh} \left[  
\max_{ i }
K \varphi \left( \frac{\rho}{N_t} \trace(\sfh {\widehat{\bQ}}^i \sfh^\dg) 
\right) \right] \nonumber 
\end{eqnarray}
where $\widehat{\bQ}^i=\frac{1}{K} \sum_{k=1}^K \bQ^i_k$ 
satisfies $\trace(\widehat{ \bQ}^i) \leq \frac{N_t N_c}{K}$ and (a) 
follows from the concavity of $\varphi(\cdot)$. Thus, the average 
mutual information with a codebook $C$ can be upper bounded by an 
appropriately generated codebook from ${\cal D}$. Since ${\cal D} \subset 
{\cal C}$, the upper bound is tight. 
\endproof

\subsection{Proof of Lemma~\ref{lem_mse}} 
\label{app_lem_mse}
From the fundamental theorem of calculus and the MMSE connection in App.~\ref{app_1}, 
we have 
\begin{eqnarray} 
\Delta I & = & \bEe_{\sfh} \left[ \int_{A  }^B \mse(x) dx   \right]
\nonumber 
\end{eqnarray}  
where $A = \frac{\rho N_c}{K} \sum_{m=1}^{N_t} \alpha_{jm} s_{im}$, 
$B = \frac{\rho N_c}{K} \lambda_{\max}(\sfh^{\dg}\sfh )$, and 
$\mse(\cdot)$ is the MSE function. Note that $A \leq B$. Since a Gaussian 
input maximizes the $\mse(\cdot)$ function~\cite{guo_mmse}, an upper bound 
to $\Delta I$ can be achieved by replacing the entropy function $\varphi(x)$ 
with that corresponding to a Gaussian input, (or equivalently, $\log(1 + x)$). 
Thus, we have 
\begin{eqnarray}
\Delta I & \leq & \bEe_{\sfh} \left[ \log(1 + B) - \log(1 + A) \right] 
\hspp = \hspp 
\bEe_{\sfh} \left[ \log \left( 1 + \frac{B-A}{1+ A} \right)   \right] 
\nonumber \\ 
& \stackrel{(a)}{\leq} & \frac{\rho N_c}{K} \cdot 
\bEe_{\sfh} \left[ \frac{\lambda_{\max}(\sfh^H \sfh) - 
\sum_{m=1}^{N_t} \alpha_{jm} s_{im} }{1+ A }  \right] 
\stackrel{(b)}{\leq} \frac{\rho N_c}{K} \cdot 
\bEe_{\sfh} \left[ \lambda_{\max}(\sfh^H \sfh) - \sum_{m=1}^{N_t} \alpha_{jm} s_{im} 
\right] \nonumber 
\end{eqnarray}
where (a) follows from log-inequality and (b) trivially. 
\endproof

\subsection{Proof of Theorem~\ref{thm_onemode_quan}} 
\label{app_onemode_quan}
We need the following proposition towards proving the theorem. 
\begin{prop} 
\label{prop_ineq_max_avg} 
For any $\{ a_{j, \hsppp k}, \hspp j=1, \hsppp \cdots, \hsppp M \}$ such that 
$\sum_{k=1}^N a_{j,k} = 1$, we have 
\beq 
\label{eq_prop_ineq_max_avg}
\max_{j = 1, \hsppp \cdots, \hsppp M} 
\sum_{k= 1}^N a_{j, \hsppp k} y_{j, \hsppp k} 
\leq \sum_{k_1 = 1}^N a_{1,\hsppp k_1} \hsppp \cdots \hsppp 
\sum_{k_{M} = 1}^N a_{M, \hsppp k_{M}} \hsppp \max\{ y_{1, \hsppp k_1}, 
\hsppp \cdots, \hsppp y_{M, \hsppp k_{M}} \}. 
\eeq
\end{prop}
\vsp 
\begin{proof}
Note that $\max \left\{y_i, \hsppp z \right\} \geq y_i$ and $\max\{y_i, \hsppp z\} 
\geq z$ for any set of real numbers $\{y_i\}_{i=1 , \hsppp \cdots , \hsppp N}$ and 
$z$. Thus, we have the inequality
\beq 
\label{ineq_max_avg}
\max \left\{ \sum_{i=1}^N \beta_i y_i, \hsppp z \right\} \leq \sum_{i=1}^N 
\beta_i \hsppp \max\{ y_i, \hsppp z\} 
\eeq
where $\sum_{i=1}^N \beta_i = 1$. 

We now prove the proposition by induction. With $M = 1$, equality holds and the 
statement is trivially valid. If the proposition holds for some $M$, we then have 
\beq
\begin{split}
\max_{j=1, \hsppp \cdots, \hsppp M+1} & 
\sum_{k=1}^N a_{j, \hsppp k} y_{j, \hsppp k} = 
\max \left  \{ \sum_{k_{M+1} =  1}^N a_{M+1,\hsppp k_{M+1}} y_{M+1,\hsppp k_{M+1}}, 
\max_{j=1, \hsppp \cdots, \hsppp M} \sum_{k=1}^N a_{j,\hsppp k} y_{j,\hsppp k} 
\right \} \nonumber \\ 
& \stackrel{(a)}{\leq}  \sum_{k_{M+1} = 1}^N a_{M+1,\hsppp k_{M+1}} \max \left 
\{ y_{M+1, \hsppp k_{M+1}}, \sum_{k_1 = 1}^N a_{1, \hsppp k_1} \hsppp \cdots \hsppp 
\sum_{k_{M}=1}^N a_{M,\hsppp k_{M}} 
\max\{ y_{1,\hsppp k_1}, \hsppp \cdots, \hsppp y_{M, \hsppp k_{M}} \} 
\right \} \nonumber \\
& \stackrel{(b)}{\leq}  \sum_{k_1=1}^N a_{1,\hsppp k_1} \hsppp \cdots \hsppp 
\sum_{k_{M} = 1}^N a_{M, \hsppp k_{M}} \sum_{k_{M+1} = 1}^N a_{M+1, \hsppp k_{M+1}} 
\hsppp 
\max \left \{ y_{M+1,\hsppp k_{M+1}},  \max\{ y_{1, \hsppp k_1}, \hsppp 
\cdots, \hsppp y_{M, \hsppp k_{M}} \} \right \} \nonumber \\
& =  \sum_{k_1 = 1}^N a_{1,\hsppp k_1} \hsppp \cdots \hsppp \sum_{k_{M} = 1}^N 
a_{M, \hsppp k_{M}} \hsppp \sum_{k_{M+1} = 1}^N a_{M+1,\hsppp k_{M+1}} \hsppp 
\max \left \{  y_{1, \hsppp k_1}, \hsppp \cdots, \hsppp  
y_{M+1, \hsppp k_{M+1}}  \right \} 
\end{split}
\eeq
where (a) follows by applying (\ref{ineq_max_avg}) on the first term in the 
max and the hypothesis in (\ref{eq_prop_ineq_max_avg}) on the second term, and (b) 
by applying (\ref{ineq_max_avg}) on the second term in the max. 
\end{proof}

\noindent {\bf \emph{ Proof of Theorem \ref{thm_onemode_quan}:}} 
To prove the theorem, we upper bound the average $\snr$ at the receiver 
\beq \nonumber
\begin{split}
\bEe & \left[  \frac{\rho N_c}{K} \max_{i,j}
\sum_{m=1}^{N_t} \alpha_{jm} s_{im}   \right] 
\stackrel{(a)} {\leq} \bEe \left[  \frac{\rho N_c}{K} \max_{j}
\sum_{m=1}^{N_t} \alpha_{jm} \max_{i} s_{im}   \right] \\
& \stackrel{(b)} {\leq} \bEe \left[  \frac{\rho N_c}{K} 
\sum_{m_1=1}^{N_t} \alpha_{1,\hsppp m_1} \hsppp \cdots \hsppp 
\sum_{m_{N_2} = 1}^{N_t} \alpha_{N_2,\hsppp m_{N_2}} \max 
\left\{ \max_{i} s_{im_1},\hsppp \cdots, \hsppp 
\max_{i} s_{im_{N_2}} \hsppp  \right\} \right] \\ 
& \stackrel{(c)} {\leq}  \bEe \left[  \frac{\rho N_c}{K} 
\max \left\{ \max_{i} s_{im_1^{\star}}, \hsppp \cdots, \hsppp 
\max_{i} s_{im_{N_2}^{\star}} \right\} \right] 
\end{split}
\eeq
where (a) and (b) follow from Prop.~\ref{prop_ineq_max_avg}, and (c) follows by 
letting 
\beq \nonumber
(m_1^{\star},\hsppp \cdots,\hsppp m_{N_2}^{\star}) 
=  \arg \max_{\{ (m_1,\hsppp \cdots, \hsppp m_{N_2}): 
\hsppp1 \leq \hsppp m_j \hsppp \leq\hsppp N_t \hsppp {\rm for} \hsppp {\rm all} \hsppp 
 j=1, \hsppp \cdots, \hsppp N_2 \} } 
\bEe \left[  
\max_j \left\{ \max_{i} s_{im_j} \right\} \right].
\eeq
The upper bound can be achieved by letting $\alpha_{jm}^{\star} = 
\delta_{m_j^{\star}m}$ which again does not depend on the channel realization. That 
is, the optimal codebook is of rank-one.
\endproof

\bibliographystyle{IEEEbib}
\bibliography{newrefs}

\end{document}